# Ultrafast Formation and Annihilation of Strongly Bound, Anisotropic Excitons

Lawson T. Lloyd*[,1], Tommaso Pincelli[1,2], Mohamed Amine Wahada[1], Alessandro De Vita[1,2], Ferdinand Menzel[3,#], Kseniia Mosina[4], Túlio H. L. G. Castro[1], Alexander Neef[1,2], Andreas V. Stier[3], Nathan P. Wilson[3,5], Zdeněk Sofer[4], Jonathan J. Finley[3], Martin Wolf[1], Laurenz Rettig[1,6], and Ralph Ernstorfer*[,1,2]

1. Department of Physical Chemistry, Fritz-Haber-Institut der Max-Planck-Gesellschaft, 14195 Berlin, Germany
2. Institut für Physik und Astronomie, Technische Universität Berlin, 10623 Berlin, Germany
3. Walter Schottky Institut and TUM School of Natural Sciences, Technische Universität München, 85748 Garching, Germany
4. Department of Inorganic Chemistry, University of Chemistry and Technology Prague, 166 28 Prague 6, Czech Republic
5. Munich Center for Quantum Science and Technology (MCQST), 80799 Munich, Germany
6. Department of Physics and Research Center OPTIMAS, RPTU University Kaiserslautern-Landau, 67653 Kaiserslautern, Rhineland-Palatinate, Germany

* Corresponding authors: lloyd@fhi-berlin.mpg.de, ernstorfer@tu-berlin.de

[#] Present Address: Dept. of Physics, University of Basel, Basel CH-4056, Switzerland

## ABSTRACT

Van der Waals (vdW) layered materials with long-range magnetic order have the potential to enable novel optoelectronic and spintronic applications. Among these, CrSBr is an air-stable, direct band gap semiconductor that hosts interlayer antiferromagnetic order, a highly anisotropic electronic structure, and strongly bound excitons. In particular, excitons in CrSBr have been shown to inherit the quasi-one-dimensional nature of the material and also couple to the underlying spin-order. However, mechanisms of exciton formation, dissociation, and interaction with free carriers remain largely unexplored, despite being crucial for spintronic and optoelectronic applications. Here, we employ time- and angle-resolved photoemission spectroscopy to map the electronic structure and excited state dynamics in CrSBr. We directly resolve an exceptionally large exciton binding energy (~800 meV) and a highly anisotropic momentum space distribution of the exciton, revealing its quasi-1D real-space character. We observe an excitation-density-dependent interconversion between bound excitons and quasi-free carriers on sub- to few-picosecond timescales, indicating that many-body effects govern the excited-state dynamics and optical properties during the initial stages of relaxation. Our work highlights the strongly bound, anisotropic character of excitons in CrSBr, as well as the microscopic interactions steering relaxation pathways after photoexcitation in elevated density regimes relevant for future device applications.

## Introduction

Two-dimensional, van der Waals (vdW) materials feature emergent properties, new degrees of freedom, and strong many-body effects that may enable next-generation devices based on engineered functionality in the atomic limit[1–3]. In 2D semiconductors such as transition-metal dichalcogenides (TMDs, *e.g.* $MoS_2$, $WSe_2$), the reduced dielectric screening and layer confinement lead to Coulomb-bound electron-hole pairs, or excitons, with binding energies of ~100s meV[4–7]. These excitons dominate the optical properties and, in turn, affect device function and performance in atomically thin optoelectronic applications[8,9]. vdW materials supporting long-range magnetic order provide additional functionality, namely for exploiting the spin degree of freedom in next-generation opto-spintronic applications built around two-dimensional vdW layers or heterostructures[10–12] and offer the opportunity to explore the interplay of many-body physics and magnetism in reduced dimensions.

Among these, CrSBr is an air-stable magnetic semiconductor hosting a direct electronic band gap, highly anisotropic optical and electronic properties, and pronounced many-body exciton physics[13–16]. Below the Néel temperature ($T_N$ ~ 132 K), CrSBr is an A-type antiferromagnet, with electron spins ferromagnetically oriented in-plane within a layer and antiferromagnetically aligned between adjacent layers[17–19]. In particular, excitons in CrSBr display pronounced magnetic order-dependent interlayer hybridization[20–26], providing a basis for the optical read-out and control of the underlying spin order in a magnetic semiconductor. These excitons have also been proposed to display a quasi-1D character with large binding energy[15,26], inheriting the structural and electronic anisotropy of the material. Previous work has observed the anisotropic conduction band dispersion and revealed rich exciton physics including, for example, exciton-polaritons[14,27,28] and exciton-magnon coupling[21–23,29]. However, the role of free carriers and the nature of exciton formation and relaxation in the early stages after photoexcitation remains largely unexplored[30,31]. Disentangling these contributions to the excited state dynamics and resulting optical and electronic properties is critical to realize new device functionalities.

Here, we employ time- and angle-resolved photoemission spectroscopy (trARPES) to directly probe the momentum-resolved excitonic signatures and photoinduced excited state dynamics in bulk CrSBr. Leveraging momentum microscopy, we image the two-dimensional momentum-space distribution of the lowest-energy exciton after photoexcitation and thereby extract the corresponding real-space exciton wavefunction, confirming its highly anisotropic nature. By resolving both the exciton and single-particle conduction band minimum simultaneously, we directly observe a large exciton binding energy ($E_b$) of ~800 meV at room temperature as well as below the Néel temperature. Performing time-resolved measurements with a wide range of excitation densities and wavelengths, we uncover an ultrafast exciton decay channel, exciton-exciton annihilation, which dominates at elevated excitation fluences and leads to a competition between bound excitons and free carriers on a sub- to few-picosecond timescale. Our results provide novel insight into the underlying physics governing the ultrafast optical response and the anisotropic, quasi-1D excitons in CrSBr. In particular, understanding the excited state dynamics and exciton physics under elevated excitation densities is also crucial for practical device applications seeking to leverage the magneto-exciton coupling of CrSBr in next-generation opto-spintronics.

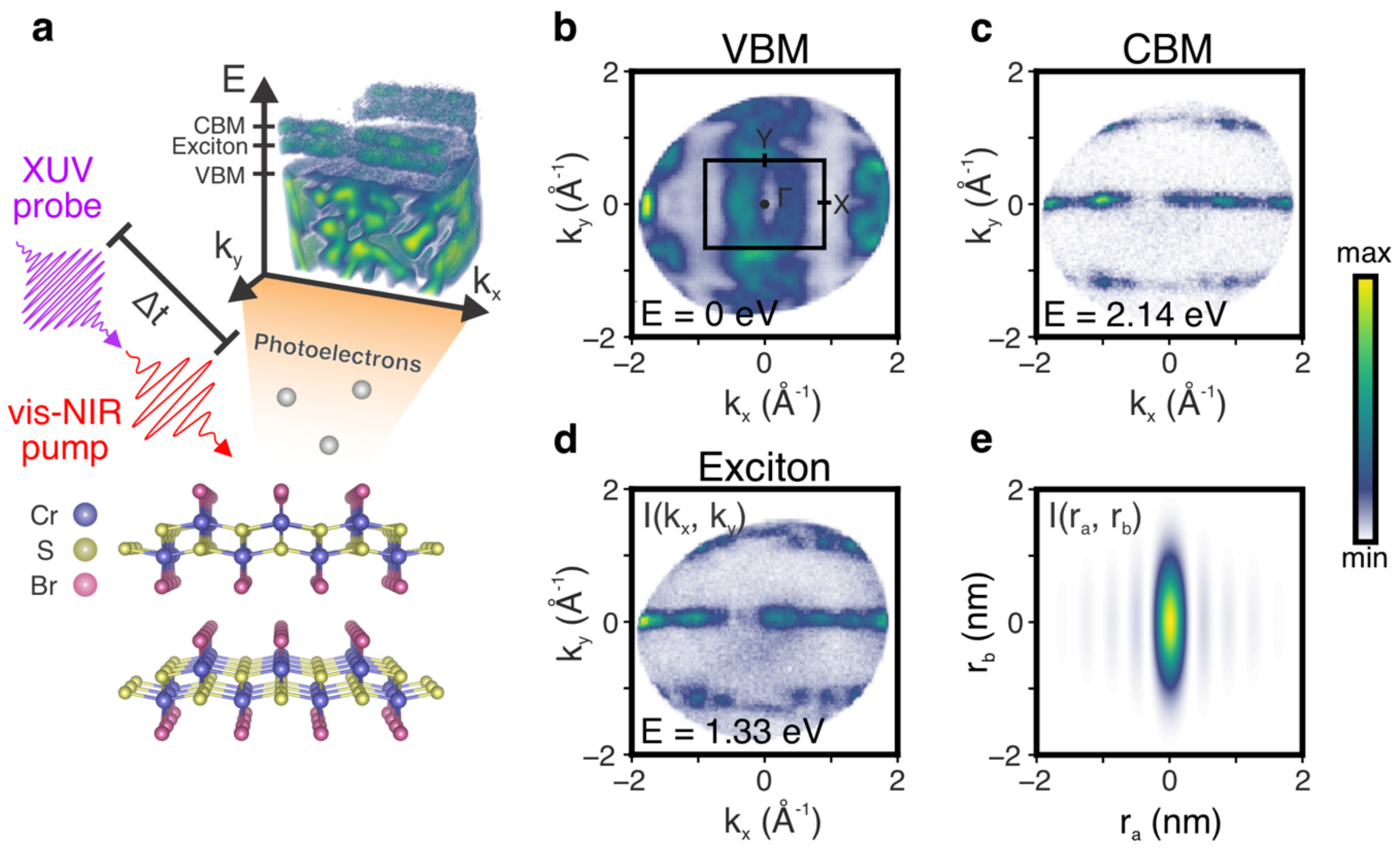

**Fig. 1: Excitons and single-particle excitations in CrSBr. a**, trARPES of bulk CrSBr probes the dynamic momentum-space occupation $I(k_x, k_y, E)$ as a function of pump-probe time delay, $\Delta t$. Momentum maps $I(k_x, k_y)$ at constant energy **b**, near the valence band maximum (VBM), and **c**, of the conduction band minimum (CBM) and **d**, of the lowest-energy exciton state after photoexcitation ($h\nu$ =1.55 eV) at T = 120 K. **e**, The real-space exciton wavefunction is retrieved by Fourier transformation of the exciton momentum map, showing highly anisotropic character with rms radii of ~0.13 nm and ~0.48 nm along the crystal *a*- and *b*-axis, respectively. E = 0 eV is referenced to the VBM at the second Γ point. Data shown are integrated over positive delay times ($\Delta t = 0$ to $\Delta t = 1000$ fs) with an energy window of 200 meV.

## Results and Discussion

### Ultrafast Momentum Microscopy

To map the electronic band structure and photoinduced dynamics of bulk CrSBr, we perform trARPES measurements utilizing a high-repetition-rate high-harmonic generation (HHG) extreme ultraviolet (XUV) laser source[32–34]. Here, tunable infrared femtosecond pump pulses polarized along the *b*-axis of the crystal, corresponding to the direction of the dipole-allowed optical transitions[20]. excite the sample and an XUV probe (~21.7 eV) pulse photoemits electrons from the sample surface after a variable time delay[35]. With momentum microscopy, we simultaneously acquire the electron binding energy in addition to both in-plane momenta, obtaining a global view of the 2D electronic band structure and exciton dynamics in a single experiment (**Fig. 1a**)[36]. In **Fig. 1b**, a two-dimensional momentum map shows the photoemission intensity $I(k_x, k_y)$ at constant energy near the valence band maximum (VBM). The central Γ point ($k_x = k_y = 0$ Å$^{-1}$) at the top of the valence band shows suppressed intensity compared to the Γ points of the second Brillouin zones at the edges of the map[37,38], which we use to define the zero energy reference ($E - E_{VBM} =$ 0 eV, **Fig. S1**).

We photoexcite ($h\nu$ =1.55 eV, ~0.9 mJ/cm$^2$, pulse duration ~40 fs, s-polarized) bulk CrSBr crystals, cleaved in ultrahigh vacuum, at T = 120 K, below $T_N$. In these conditions, we excite above the lowest-lying bright exciton resonance[39,40] and estimate an exciton density of ~3×10$^{13}$/cm$^2$ or ~0.05 excitations per unit cell (see the **Methods**). After photoexcitation, we observe two distinct features around E ~ 1.3 eV and E ~ 2.1 eV relative to the VBM, each exhibiting an anisotropic momentum space distribution and quasi-flat dispersion along $\Gamma - X$ at low temperature (**Fig. 1a,c,d**) as well as at room temperature (**Fig. 2a**). Such an anisotropic dispersion of the conduction band has been predicted theoretically[15,20] and recently observed in equilibrium ARPES of few-layer CrSBr in contact with metallic substrates[38] and after alkali metal dosing of bulk CrSBr[41].

While initial experimental studies estimated the electronic band gap of CrSBr to be around 1.5 eV [17], recent theoretical and experimental ARPES works[37,41–43] have suggested a larger band gap of roughly 2 eV. Optical spectroscopy measurements have shown the lowest-energy exciton resonance at around ~1.35 eV. Therefore, we assign these two features to the lowest-energy bound exciton state and the single-particle conduction band minimum (CBM) of CrSBr, respectively. To more precisely determine the energies of these two features observed in our trARPES measurements, we fit the momentum-integrated energy distribution curves (EDCs) to a sum of two Gaussian functions (**Figure 2a**), obtaining a peak energies of 1.25 eV and 2.04 eV for the exciton and CBM, respectively, at room temperature. Our assignment of the exciton and CBM is further supported by time-resolved experiments with pump photon energies above the band gap, as discussed below where we also expand on the origin of the CBM after pumping below the band gap. Therefore, we confirm the flat dispersion of the conduction band along $\Gamma - X$ and experimentally reveal the theoretically predicted quasi-1D character of the exciton. Additional momentum maps and trARPES spectra are displayed in **Figures S2-S6**.

By resolving both the exciton and CBM signatures in a single measurement, we directly determine the exciton binding energy from the energy difference of these photoemission features[44,45], extracting an exceedingly large exciton binding energy of $E_b = 792 \pm 4$ meV at room temperature (**Figs. 2a, S7**). Momentum-dependence of the exciton binding energy is displayed in **Figure S8**. Below the Néel temperature at T = 120 K, the band gap slightly increases, with the exciton and CBM energies found to be 1.34 eV and 2.14 eV, respectively (**Figure S9**). Here, we observe a similar binding energy of $E_b = 793 \pm 11$ meV. A recent work reported an increased exciton binding energy (~20 meV) in the antiferromagnetic phase below $T_N$ when excitons are confined within a single layer, leading to a surface-confined exciton[46,47]. We note that due to the surface selectivity of ARPES experiments, excitons probed on the surface of a bulk material experience reduced screening compared to those in the interior. Our measurement of the exciton binding energy in CrSBr is nonetheless almost one order of magnitude larger than what has been previously observed in analogous experiments on other bulk 2D vdW systems, such as the A exciton in bulk TMDs[45]. Therefore, the high exciton binding energy (~800 meV) that we observe is reflective of the intrinsic properties of CrSBr.

By resolving both in-plane momenta simultaneously, momentum microscopy provides powerful insights into the orbital and real-space character of excitons in many molecular and inorganic semiconductor systems[36,48]. Excitons in CrSBr are predicted to be spatially anisotropic with a ratio of ~1:3 between the radius of the exciton along the crystal *a*- and *b*-axes with characters in between ideal limits of site-localized Frenkel and delocalized Wannier-Mott characters[20,15,49,50]. Leveraging the two-dimensional momentum space information, we retrieve the real-space wavefunction envelope of the lowest-energy exciton in CrSBr by Fourier transformation of the momentum maps corresponding to the exciton[45,51] (see the **Supplementary Information**). Following its anisotropic momentum distribution (**Fig. 1d**), the exciton wavefunction displays an elongated distribution in real-space extended along the b-axis of the crystal (**Fig. 1e**) and localized along the weakly coupled, quasi-1D Cr-S chains[15,41]. From this, we determine the exciton's root mean square (rms) radii (**Fig. S10**) as ~0.13 nm and ~0.55 nm along the crystal a- and b-axis, respectively. At room temperature (**Fig. S11**), excitons are even more strongly localized along the b-axis (~0.48 nm) and delocalized between vdW layers. These values support a predominantly Frenkel-like character of the lower lying exciton[26] and are consistent with the large binding energy we observe, as well as recent estimates[52]. Our observations provide direct confirmation of the strongly bound ($E_b$ ~ 800 meV) and quasi-1D nature of the exciton in CrSBr.

The observation of the CBM feature at E ~ 2 eV is striking, considering the pump excitation energy of 1.55 eV is well below our assignment of the electronic band gap. **Figure 2** shows the room-temperature dispersion along $\Gamma - \mathrm{X}$ as well as the ultrafast dynamics of the exciton and CBM features at Γ. While there is a rapid rise of the exciton signal near zero time delay, the CBM exhibits a delayed rise after photoexcitation, pointing to a dynamical process that populates this state on a few hundreds of femtosecond timescale.

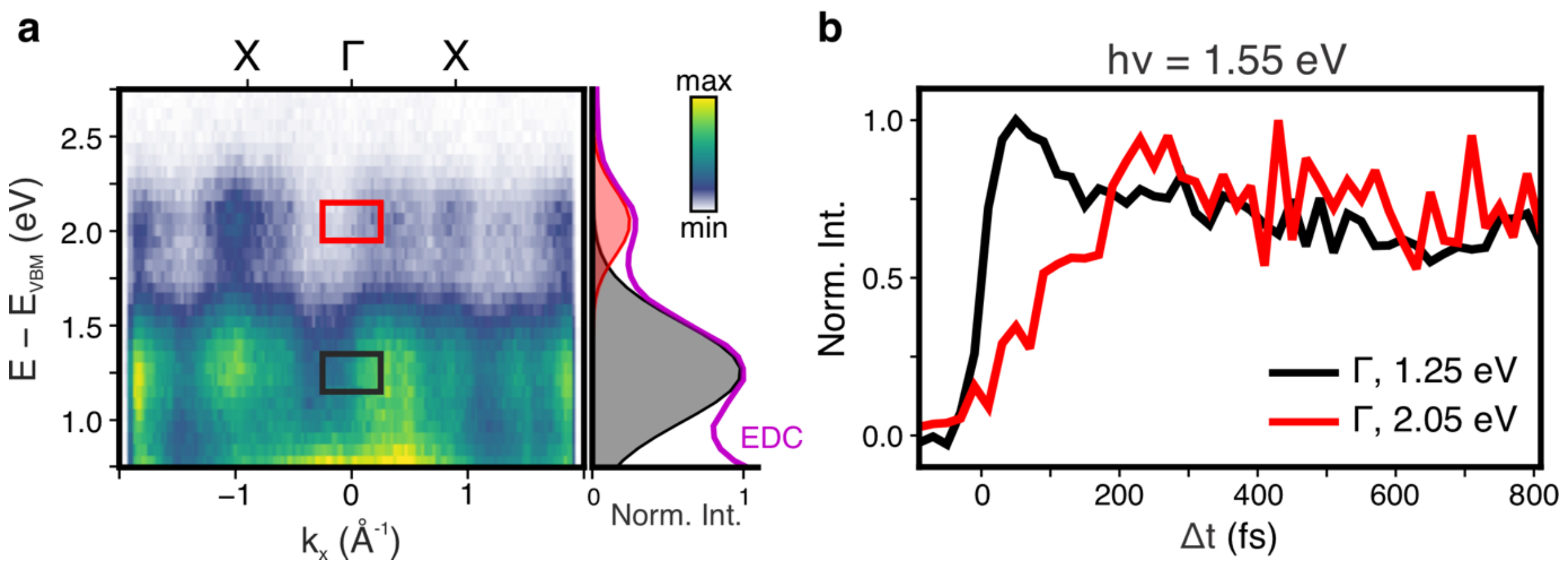

**Fig. 2: Anisotropic dispersion and ultrafast dynamics in CrSBr. a**, Delay-integrated excited state signal along $\Gamma - \mathrm{X}$ ($k_y = 0$ Å$^{-1}$) showing the flat dispersion of both the conduction band minimum (CBM) and the exciton. The energetic assignment of these states is extracted from a fit to the momentum-integrated EDC (violet) and the energy difference corresponds to the exciton binding energy, $E_b$ ~ 800 meV. **b**, Time traces at Γ show that the CBM (red) exhibits a delayed rise compared to the exciton (black) after photoexcitation ($h\nu$ =1.55 eV). The time traces are both individually normalized and integrated in an energy and momentum window of 200 meV and 0.5 Å$^{-1}$ in both $k_x$ and $k_y$, respectively. Data is acquired at T = 300 K.

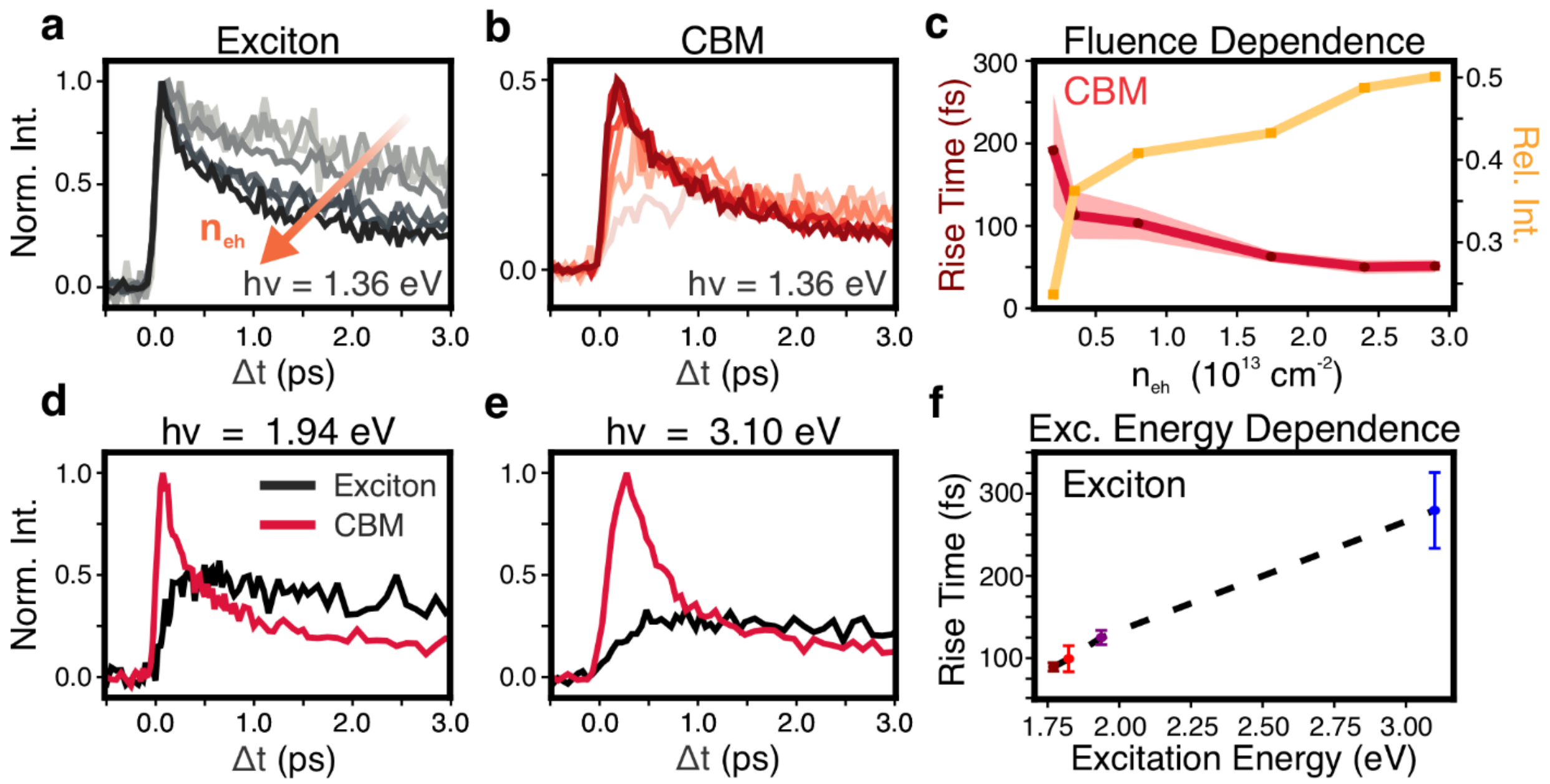

**Fig. 3: Pump excitation energy and fluence-dependent dynamics.** Dynamics for the **a** exciton and **b** CBM features integrated along $\Gamma - X$ for near-resonant exciton excitation ($h\upsilon$ =1.36 eV) with varying excitation densities ($n_{eh}$) between 0.2 – 2.9 × $10^{13}$ $cm^{-2}$. **c**, Fluence-dependent rise time and relative maximum intensity of the CBM to the exciton feature. Dynamics of the exciton (black) and CBM (red) features for **d** $h\upsilon$ =1.94 eV and **e** $h\upsilon$ =3.10 eV. **f,** Excitation energy-dependent rise time of the exciton signal. The time traces are integrated over 0.1 eV and normalized to the respective exciton (**a, b**) or CBM (**d, e**) signal after a constant background subtraction of the negative time delay signal.

## Time-resolved measurements with varying excitation density and photon energy

To better understand the dynamics leading to the population of CBM after photoexcitation, we perform further trARPES measurements with varying excitation fluence and energy at room temperature. For these data, we integrate the signal along $\Gamma - X$ and subtract a constant background due to the presence of a finite signal above the VBM overlapping with the exciton and CBM features (**Figs. S12, S13**). **Figure 3a,b** shows dynamics of the exciton and CBM features integrated along $\Gamma - X$ with excitation close to the exciton resonance ($h\upsilon$ = 1.36 eV, ~50 fs) for varying incident excitation fluences ($n_{eh}$) between ~ 0.1 – 1.5 mJ/$cm^2$. Under these conditions, we estimate photoexcited exciton densities of $n_{eh}$ ~ 0.2 – 3 × $10^{13}$ $cm^{-2}$ (see **Methods**). Similar to the 1.55 eV excitation, we observe a delayed rise of the CBM signal relative to the exciton. Interestingly, we also observe a strong dependence on excitation fluence in both the initial decay dynamics of the exciton signal (**Fig. 3a**) and the rise time and relative intensity of the CBM feature (**Fig. 3b,c**). As the excitation density increases, the exciton signal features an increasingly faster decay component in the first few hundred femtoseconds. At the same time, the CBM signal rises more quickly, from a ~200 fs to ~ 50 fs timescale, with the relative CBM signal intensity approaching ~50% of the exciton.

For excitation with increasingly higher photon energies, we observe a reversal in the early-time dynamics and relative intensities of the CBM and exciton such that the CBM state is transiently populated first (**Fig. 3d,e**), followed by a rise of the exciton signature on a few-hundred femtosecond timescale (**Figs. 3f, S14**). This observation is most prominent with 3.10 eV excitation, which populates electrons with excess energy far above the conduction band minimum and in which excitons form within a ~300 fs timescale. Temperature-dependent dynamics on similar few-hundred femtosecond timescales have been attributed to electron-phonon coupling and carrier cooling after above-gap excitation[31]. Notably, the exciton formation and CBM dynamics display minimal excitation fluence dependence (**Fig. S15**), which may indicate that excitons form partially through a geminate process directly from photoexcitation, as opposed to from thermalized free carriers at the band edge[53–55]. After the initial hot electron relaxation dynamics to the CBM, bound excitons and conduction band electrons at the CBM coexist. Such a dynamic and robust exciton formation in CrSBr is likely due to the very high exciton binding energy that we observe in this material, similar to exciton formation in monolayer TMDs[56,53]. For the excitation energies used in this study, 1.94 eV excitation features the sharpest rise of the CBM population, consistent with our conduction band assignment near ~2 eV as well as recent estimates of the electronic band gap by previous ARPES measurements[37,42,41]. We note that a second excitonic signature with high oscillator strength at 1.77 eV has also been reported[57], in addition to a prominent optically dark exciton at 1.46 eV [58–60]. Additional data below $T_N$ is (**Fig. S16**) exhibits similar behavior.

The fluence-dependent decay dynamics after near-resonant exciton excitation (**Fig. 3a,b**) suggest that a multi-exciton scattering mechanism is responsible for the sub-picosecond dynamics under these excitation densities. At elevated quasiparticle densities, the electron-hole Coulomb interaction can become screened, reducing the binding energy and leading to a renormalization of the electronic bandgap, termed band gap renormalization (BGR)[61,62]. At a critical threshold, $N_M$, the exciton gas can undergo a phase transition to an ionized plasma of quasi-free carriers, referred to as the excitonic Mott transition. The nature of the Mott transition has been a subject of intense research[61,63–65] and is critical to understand semiconductor materials such as 2D TMDs, whose excitons largely determine the optical response and electronic properties[8]. Considering the extracted effective exciton radius, we estimate the Mott threshold as $N_M = 1/a_{eff}^2 \approx 4 \times 10^{14}$, an order of magnitude greater than the fluences used in these experiments. Indeed, we observe a reduction of the exciton binding energy of only 58 meV $\pm$ 4 meV at our highest fluence, indicating bound excitons remain stable in this regime (**Figs. S21,S22**).

Based on these considerations, an excitonic Mott transition is unlikely to be the primary driving force for the observed features and dynamics we observe. However, a separate mechanism, namely exciton-exciton annihilation (EEA), may also occur at elevated excitation densities below the Mott transition[66,67]. EEA is an Auger-type, two-exciton nonradiative decay pathway and acts as a prominent loss channel for excitonic materials at high excitation densities. EEA has been widely studied in nanomaterials including TMDs[68–70], and also in quasi-1D anisotropic systems such as carbon nanotubes[71–73] and black phosphorus[74]. During EEA, one exciton non-radiatively recombines and leads to the dissociation of another exciton, promoting its constituent carriers higher and lower into the single-particle conduction and valence bands with excess energy, respectively. Experimentally, EEA manifests as second-order kinetics with respect to the excitation density, with an increasingly faster decay component dominating with elevated excitation fluence.

**Global fitting the dynamics to an EEA-based rate-equation model**

To gain further insight into the observed dynamics and test the role of EEA and exciton formation in CrSBr, we employ a coupled rate-equation model (**Eqs. 1-3**) considering populations of excitons ($N_X$), conduction band electrons at the CBM ($N_{CBM}$), and hot carriers high above the CBM ($N_{hc}$) [75]. Exciton recombination ($\tau_r$), two-body exciton-exciton annihilation ($\gamma_{EEA}$), exciton formation from free carriers ($\tau_f$), and hot carrier relaxation ($\tau_{hc}$) processes after photoexcitation are illustrated in **Fig. 4a**. The constant $F_i$ ($H_i$) is proportional to the fluence and is used to scale the excitation density in the case of exciton (above-gap) excitation, which is modelled as a Gaussian source term. We globally fit the exciton and conduction band electron dynamics for a range of excitation fluences to extract the above time and rate constants, considering both near-resonant exciton (1.36 eV) and above-gap (3.10 eV) excitation (**Figs. 4b-e,S19**). We include a free parameter to correct the relative amplitudes of each trace. The exciton recombination time constant is fixed at $\tau_r$ = 20 ps based on previous reports[30] and data acquired at longer delay times (**Fig. S18**).

$$\frac{dN_X}{dt} = F_i \times G - \frac{N_X}{\tau_r} + \frac{N_{CBM}}{\tau_f} - \gamma_{EEA} {N_X}^2 \tag{1}$$

$$\frac{dN_{CBM}}{dt} = \frac{N_{hc}}{\tau_{hc}} - \frac{N_{CBM}}{\tau_f} + \gamma_{EEA} \frac{{N_X}^2}{2} \tag{2}$$

$$\frac{dN_{hc}}{dt} = H_i \times G - \frac{N_{hc}}{\tau_{hc}} \tag{3}$$

Overall, the EEA-based global model reproduces our observations well, including the evolution of the rise time of the CBM and the faster initial decay dynamics of the exciton with increasing excitation fluence in the case of near-resonant exciton excitation and the reversal when considering above-gap excitation. We extract time constants of $\tau_f = 109 \pm 4$ fs and $\tau_{hc} = 383 \pm 12$ fs for exciton formation and hot carrier cooling, respectively. The retrieved exciton formation time $\tau_f$ is consistent with exciton formation times when considering different excitation energies near the CBM (**Fig. S14**). Note that in this model, we consider only exciton formation as a decay pathway for free carriers after either formation *via* EEA or hot carrier relaxation to the CBM. We determine an EEA rate of $\gamma_{EEA} = 0.090 \pm 0.002$ $cm^2/s$, corresponding to EEA timescales of ~5.5 ps to ~370 fs for the excitation fluences used in our measurements (0.2 – 3 × $10^{13}$ $cm^{-2}$). This EEA rate is similar to rates observed in monolayer TMDs such as exfoliated $MoS_2$ (0.043 $cm^2/s$) [69] and $WS_2$ grown by chemical vapor deposition (0.089 $cm^2/s$) [76], but substantially higher than 2D quantum well systems (~$10^{-3}$ $cm^2/s$) [70,77]. Defects have been suggested to play a significant role in the nonradiative losses and exciton recombination dynamics, contributing to the observed annihilation rates.[68] We observe a static in-gap signal at equilibrium indicative of n-type doping (**Figure S9**).[78] An ultrafast defect-assisted exciton annihilation mechanism, as has also previously been in TMDs[79,80], may therefore also be relevant in CrSBr. Strong exciton-phonon coupling in CrSBr[81] may also lead to enhanced phonon-assisted Auger-type decay pathways[82,83]. Nevertheless, this EEA process therefore acts as a prominent nonradiative decay pathway for photoexcited excitons at elevated excitation densities, leading to a rapid decay of the total excited state population (**Fig. S17**). Together, the processes of ultrafast EEA and robust exciton formation lead to a strong competition and interconversion between quasi-free conduction band electrons and bound excitons on few hundred femtosecond timescales.

Sequential absorption of photons within the pump pulse may also be responsible for the transient CBM population after near-resonant exciton excitation, as previously reported for $ReSe_2$[84].Exciton photoionization by sequential absorption is expected to be population conserving. However, we observe a stark fluence-dependent dynamics in the total excited state population persisting after the pulse duration (~50 fs), which is not reproduced by a model considering sequential absorption in place of EEA (**Fig. S20**), suggesting it is not the sole origin of the dynamics observed in our experiments.

We note that this EEA-based rate-equation model neglects effects originating from, for example, coherent electronic polarization dynamics or many-body effects which may be especially relevant at short time delays during, during and directly after optical excitation.[85] In addition, our model does not capture the complex correlations that arise from interactions with other degrees of freedom, such as the spin or lattice through coupling to magnons or phonons, and which influence the optical properties and excited state dynamics, especially in the magnetically ordered phase.

In CrSBr, the quasi-1D character may modify the exciton dynamics and EEA process. In a confined 1D system, excitons interact via their nearest neighbors, leading to a time-dependent EEA rate.[86] Excitons in CrSBr inherit the quasi-1D character and material anisotropy and are confined along the crystal b-axis. However, at elevated excitation densities the effect of this confinement on the EEA behavior may be reduced, as observed for black phosphorus[74], as the inter-exciton distance is reduced and interactions along the a-axis are non-negligible.

In this work, we provide for the first time a direct, momentum-resolved view of the strongly bound, anisotropic excitons in the bulk van der Waals magnetic semiconductor CrSBr. Using time-resolved momentum microscopy, we have extracted an exceedingly large exciton binding energy of ~800 meV for excitons probed on the surface of bulk CrSBr, and we retrieved their anisotropic real-space excitonic distribution, with radii of ~0.13 nm and ~0.55 nm along the crystalline a- and b-axes at T = 120 K. These results support a picture of excitons confined along the quasi-1D Cr-S chains and extending along the crystal b-axis, with a weak dependence of key exciton properties on the magnetic state of the material. Upon near-resonant excitation of the lowest-energy exciton, a many-body decay channel, namely exciton-exciton annihilation, dominates with increasing excitation density above $n_{eh} \sim 10^{13}$ carriers/cm$^2$, leading to the generation of quasi-free carriers and a rapid decay of the total excited state population on timescales of sub- to few-picoseconds. With substantially higher, above-gap excitation of quasi-free electrons, we observe the relaxation of hot electrons to the band edge before the formation of strongly bound excitons, which appear extremely robust in this system. The competition between the annihilation and formation of excitons drives the femtosecond and picosecond excited state dynamics after optical excitation in CrSBr, dominating over excitation-induced weakening of the exciton binding energy in described by the excitonic Mott transition. Our work provides important insight into the many-body interactions and physics underlying the strongly bound, quasi-1D excitons in CrSBr, critical for understanding and steering functionalities in next-generation opto-spintronic devices.

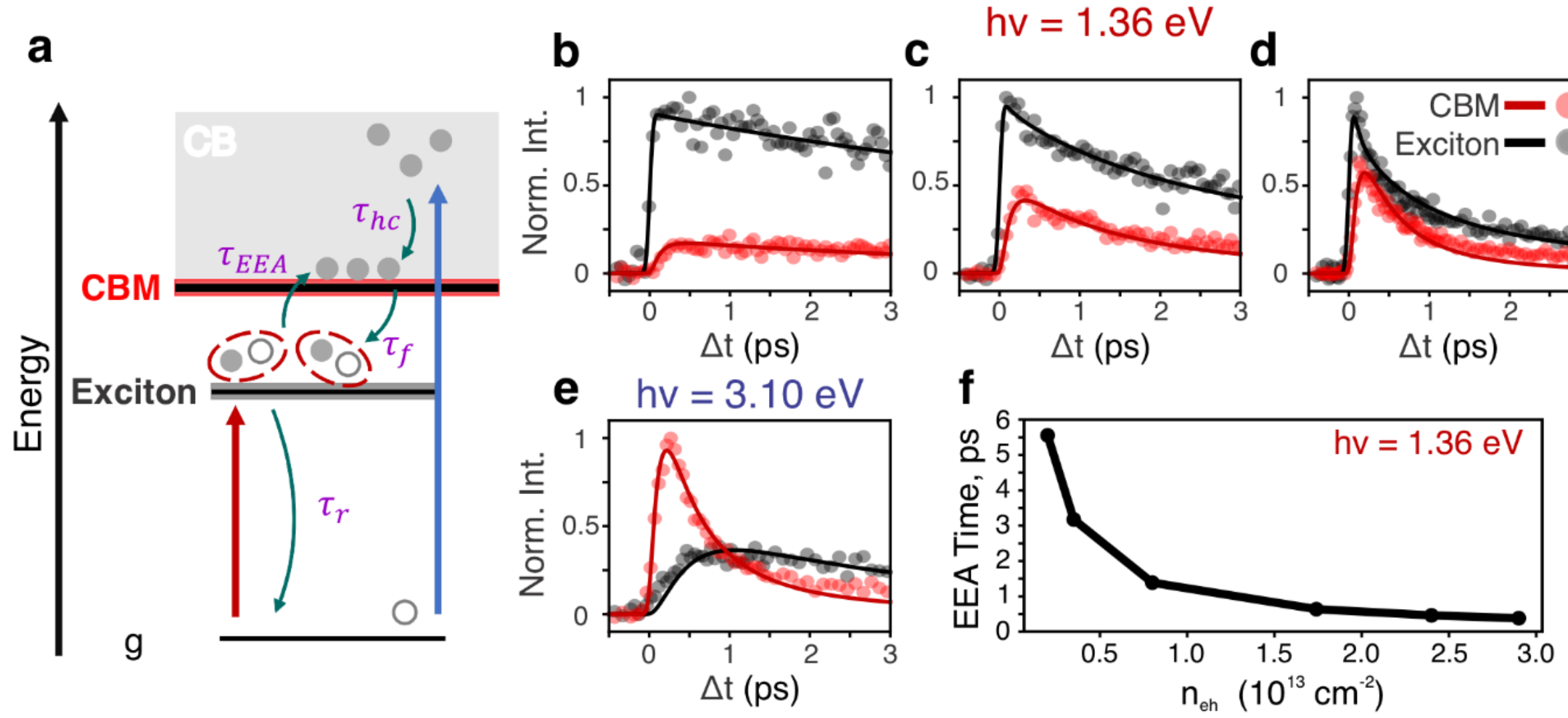

**Fig. 4: Exciton and free carrier interconversion. a**, After photoexcitation near the exciton resonance, exciton-exciton annihilation leads to quasi-free electrons near the CBM, while above-bandgap excitation populates hot carriers with excess energy which then relax to form excitons. **b – e**, Global fitting of the data, considering both 1.36 eV and 3.10 eV excitation, to a coupled rate-equation model reproduces the fluence- and excitation energy-dependent trends. **f**, Effective EEA timescales at different excitation fluences extracted from the EEA coefficient. The excitation density for 1.36 eV excitation (**b – d**) is 0.2, 0.8, and 2.9 × $10^{13}$ cm$^{-2}$, respectively.

## Methods

CrSBr synthesis

CrSBr bulk crystal was synthesized by a chemical vapor transport method. Elemental chromium (99.99%, -60 mesh, Chemsavers, USA), sulfur (99.9999%, 1 - 6 mm Wuhan Xinrong New Materials Co., China), and bromine (99.9999%, Merck, Czech Republic) were placed with a stoichiometry of 1:1:1 in a quartz ampoule (60x250mm). The amounts corresponded to 30g of CrSBr using 2 at.% excess of Br and S. The ampoule, cooled with liquid nitrogen, was melt sealed under high vacuum using an oil diffusion pump with liquid nitrogen trap. The ampoule was then heated in a crucible furnace at 400 °C for 50 hours, at 500°C for 50 hours and at 600°C for 50 hours. The top part of the ampoule was kept under 100 °C to avoid excessive pressure formation in the ampoule. The reacted ampoule was subsequently placed in a two zone furnace, where first the source zone was heated to 700 °C and the growth zone to 900 °C. After two days the thermal gradient was reversed and the source zone was kept at 900 °C and growth zone at 800 °C for 10 days. Finally, the ampoule was cooled to room temperature and opened inside an argon-filled glovebox.

ARPES Measurements

ARPES measurements were performed using a home-built, tabletop high-harmonic generation (HHG) source operating at ~500 kHz and delivering probe pulses at ~21.7 eV for photoemission[32,33]. For momentum-microscopy-based trARPES experiments, the output of an optical parametric chirped pulse amplification (OPCPA) was used to drive HHG by tight focusing of its second harmonic (400 nm) in an argon gas jet in vacuum. A portion of the OPCPA output (800 nm, ~40 fs) was used for pump pulses in time-resolved momentum microscopy experiments with 1.55 eV excitation. For other trARPES experiments, HHG was driven with 515 nm from the second harmonic of a portion of the 1030 nm output of a frontend laser (Carbide, Light Conversion). This 1030 nm output was also used to pump an OPA (Orpheus, Light Conversion) for tunable pump pules (640-940 nm, ~50 fs). With the OPCPA-based setup, the driving beam (400 nm) was separated from the generated XUV by an aperture in the far-field while in the second case, the XUV was isolated from the driving beam (515 nm) by reflection on a SiC substrate at Brewster's angle. The HHG order of interest near ~21.7 eV was isolated by a multilayer XUV mirror focusing the XUV onto the sample, as well as a 400-nm thick transmissive Sn filter. Before the measurements, bulk CrSBr crystals were mechanically cleaved in ultrahigh vacuum in base pressures better than ~$10^{-10}$ mbar. A six-axis manipulator (Carving, SPECS GmbH) is used to control the alignment of the sample to either a time-of-flight momentum microscope (METIS1000, SPECS GmbH) or a hemispherical electron energy analyzer (PHOIBOS150, SPECS GmbH)[34]. In all measurements, the XUV pulses were p-polarized while the pump pulses were s-polarized. The excitation density is calculated[87] considering the sample dielectric function[15] and pump wavelength, polarization, angle of incidence, and spot size (~110 $\mu$m full-width at half-maximum) on the sample. In our measurements, we estimate the temporal resolution from the pump-probe cross correlation to be ~40 fs (400 nm HHG driver) and ~80 fs (515 nm HHG driver). The experimental energy resolution is ~120 meV (400 nm HHG driver) and ~40 meV (515 nm HHG driver), dominated by the bandwidth of the XUV probe pulses.

## Data Availability

The datasets generated and analyzed in this study are made available on a public data repository:

## Acknowledgements

This work was supported by the Deutsche Forschungsgemeinschaft (DFG) within the Transregio TRR 227 Ultrafast Spin Dynamics (Project B07) and through the priority program SPP2244 (Project No. 443366970), the European Research Council (ERC) under the European Union's Horizon 2020 research and innovation program (H2020-FETOPEN-2018-2019-2020-01, OPTOLogic – Grant Agreement No. 899794), and the Max Planck Society. L.T.L. acknowledges financial support from the Alexander von Humboldt Foundation. A.D.V. acknowledges support from BERLIN QUANTUM, an initiative endowed by the Innovation Promotion Fund of the city of Berlin. N.P.W., A.V.S. and J.J.F. acknowledge the German Science Foundation (DFG) for financial support via the Priority Programmes SPP 2244 "2D Materials Physics of van der Waals heterobilayer", as well as the clusters of excellence MCQST (EXS-2111) and e-conversion (EXS-2089). Z.S. was supported by project LUAUS25268 from Ministry of Education, Youth and Sports (MEYS) and by the project Advanced Functional Nanorobots (reg. No. CZ.02.1.01/0.0/0.0/15_003/0000444 financed by the EFRR). The authors thank Florian Dirnberger and Philip Hofmann for fruitful discussions and Tim Sanetra for assistance in preliminary analysis. The CrSBr visualization in **Fig. 1** was produced using the VESTA software[88].

## Author Contributions

R.E., A.V.S., T.P. and L.T.L. initiated the project. L.T.L., T.P., M.A.W, A.D.V., T.H.L.G.C., and A.N. performed ARPES experiments. F.M., N.P.W., K.M., and Z.S. provided bulk crystal samples. M.W., R.E., L.R., and J.J.F provided supervision and experimental infrastructure. L.T.L. analyzed the data and wrote the manuscript with input from all authors.

## Competing Interests

The authors declare no competing interests.

**Fig. 1: Excitons and single-particle excitations in CrSBr. a**, trARPES of bulk CrSBr probes the dynamic momentum-space occupation $I(k_x, k_y, E)$ as a function of pump-probe time delay, $\Delta t$. Momentum maps $I(k_x, k_y)$ at constant energy **b**, near the valence band maximum (VBM), and **c**, of the conduction band minimum (CBM) and **d**, of the lowest-energy exciton state after photoexcitation ($h\upsilon$ =1.55 eV) at T = 120 K. **e**, The real-space exciton wavefunction is retrieved by Fourier transformation of the exciton momentum map, showing highly anisotropic character with rms radii of ~0.13 nm and ~0.48 nm along the crystal *a*- and *b*-axis, respectively. E = 0 eV is referenced to the VBM at the second Γ point. Data shown are integrated over positive delay times ($\Delta t = 0$ to $\Delta t = 1000$ fs) with an energy window of 200 meV.

**Fig. 2: Anisotropic dispersion and ultrafast dynamics in CrSBr. a**, Delay-integrated excited state signal along $\Gamma - \mathrm{X}$ ($k_y = 0$ Å$^{-1}$) showing the flat dispersion of both the conduction band minimum (CBM) and the exciton. The energetic assignment of these states is extracted from a fit to the momentum-integrated EDC (violet) and the energy difference corresponds to the exciton binding energy, $E_b \sim 800$ meV. **b**, Time traces at Γ show that the CBM (red) exhibits a delayed rise compared to the exciton (black) after photoexcitation ($h\upsilon$ =1.55 eV). The time traces are both individually normalized and integrated in an energy and momentum window of 200 meV and 0.5 Å$^{-1}$ in both $k_x$ and $k_y$, respectively. Data is acquired at T = 300 K.

**Fig. 3: Pump excitation energy and fluence-dependent dynamics.** Dynamics for the **a** exciton and **b** CBM features integrated along $\Gamma - \mathrm{X}$ for near-resonant exciton excitation ($h\upsilon$ =1.36 eV) with varying excitation densities ($n_{eh}$) between 0.2 – 2.9 × $10^{13}$ cm$^{-2}$. **c**, Fluence-dependent rise time and relative maximum intensity of the CBM to the exciton feature. Dynamics of the exciton (black) and CBM (red) features for **d** $h\upsilon$ =1.94 eV and **e** $h\upsilon$ =3.10 eV. **f,** Excitation energy-dependent rise time of the exciton signal. The time traces are integrated over 0.1 eV and normalized to the respective exciton (**a, b**) or CBM (**d, e**) signal after a constant background subtraction of the negative time delay signal.

**Fig. 4: Exciton and free carrier interconversion. a**, After photoexcitation near the exciton resonance, exciton-exciton annihilation leads to quasi-free electrons near the CBM, while above-bandgap excitation populates hot carriers with excess energy which then relax to form excitons. **b – e**, Global fitting of the data, considering both 1.36 eV and 3.10 eV excitation, to a coupled rate-equation model reproduces the fluence- and excitation energy-dependent trends. **f**, Effective EEA timescales at different excitation fluences extracted from the EEA coefficient. The excitation density for 1.36 eV excitation (**b – d**) is 0.2, 0.8, and 2.9 × $10^{13}$ cm$^{-2}$, respectively.